\documentclass[utf8]{frontiersSCNS} 

\usepackage{url,hyperref,lineno,microtype,subcaption}
\usepackage[onehalfspacing]{setspace}
\usepackage{caption}

\def\keyFont{\fontsize{8}{11}\helveticabold }
\def\firstAuthorLast{Panda {et~al.}} 
\def\Authors{Swayamtrupta Panda\,$^{1,2,*}$, Bo{\.z}ena Czerny\,$^{1,2}$ and Conor Wildy\,$^{1}$}
\def\Address{$^{1}$Center for Theoretical Physics, Polish Academy of Sciences, Al. Lotnikow 32/46, 02-668 Warsaw, Poland \\
$^{2}$Nicolaus Copernicus Astronomical Center, Polish Academy of Sciences, Bartycka 18, 00-716 Warsaw, Poland}
\def\corrAuthor{Swayamtrupta Panda}

\def\corrEmail{panda@cft.edu.pl}

\begin{document}
\onecolumn
\firstpage{1}

\title[The physical driver of the optical Eigenvector 1 in Quasar Main Sequence]{The physical driver of the optical Eigenvector 1 in Quasar Main Sequence} 

\author[\firstAuthorLast ]{\Authors} 
\address{} 
\correspondance{} 

\extraAuth{}

\maketitle

\begin{abstract}

\section{}


Quasars are complex sources, characterized by broad band spectra from radio through optical to X-ray band, with numerous emission and absorption features. This complexity leads to rich diagnostics. However, \citet{bg92} used Principal Component Analysis (PCA), and with this analysis they were able to show significant correlations between the measured parameters. The leading component, related to Eigenvector 1 (EV1) was dominated by the anticorrelation between the Fe${\mathrm{II}}$ optical emission and [OIII] line and EV1 alone contained 30\% of the total variance. It opened a way in defining a quasar main sequence, in close analogy to the stellar main sequence on the Hertzsprung-Russel (HR) diagram (\citealt{sul01}). The question still remains which of the basic theoretically motivated parameters of an active nucleus (Eddington ratio, black hole mass, accretion rate, spin, and viewing angle) is the main driver behind the EV1. Here we limit ourselves to the optical waveband, and concentrate on theoretical modelling the Fe${\mathrm{II}}$ to H$\mathrm{\beta}$ ratio, and we test the hypothesis that the physical driver of EV1 is the maximum of the accretion disk temperature, reflected in the shape of the spectral energy distribution (SED). We performed computations of the  H$\mathrm{\beta}$ and optical Fe${\mathrm{II}}$ for a broad range of SED peak position using CLOUDY photoionisation code. We assumed that both H$\mathrm{\beta}$ and Fe${\mathrm{II}}$ emission come from the Broad Line Region represented as a constant density cloud in a plane-parallel geometry. We expected that a hotter disk continuum will lead to more efficient production of Fe${\mathrm{II}}$ but our computations show that the Fe${\mathrm{II}}$ to  H$\mathrm{\beta}$ ratio actually drops with the rise of the disk temperature. Thus either hypothesis is incorrect, or approximations used in our paper for the description of the line emissivity is inadequate.

\tiny
 \keyFont{ \section{Keywords:} quasars: broad line region, Eigenvector 1,  FeII strength, accretion disk temperature, constant density, photoionisation: CLOUDY} 
\end{abstract}

\section{Introduction}


Quasars are rapidly accreting supermassive black holes at the centres of massive galaxies. In type 1 AGN, we see the nucleus directly, the continuum emission dominating the energy output in the optical/UV band comes from an accretion disk surrounding a supermassive black hole \cite[e.g.][]{Czerny1987,cap2015}, and the optical/UV emission broad emission lines, Fe${\mathrm{II}}$ pseudo-continuum and Balmer Component are usually considered to be coming from the Broad Line Region (BLR) clouds. Broad band spectral properties and line emissivity are highly correlated \cite{bg92,sul00,sul02,sul07,yip04,sh14,sun15}, and the PCA analysis is a powerful tool herein. As suggested by \cite{sul01}, those correlations allow the identification of the quasar main sequence, analogous to the stellar main sequence on the HR diagram where the classification was also based purely on spectral properties of the stellar atmospheres. The stellar main sequence found the dependence of the spectra on the effective temperature of stars. Quasar main sequence was suggested to be driven mostly by the Eddington ratio \cite{bg92,sul00,sh14} but also on the additional effect of the black hole mass, viewing angle and the intrinsic absorption \cite{sh14,sul00,kura09}.
\par 
We postulate that the true driver behind the R$_{\mathrm{FeII}}$ is the maximum of the temperature in a multicolor accretion disk which is also the basic parameter determining the broad band shape of the quasar continuum emission. The hypothesis seems natural because the spectral shape determines both broad band spectral indices as well as emission line ratios, and has already been suggested by \cite{b07}. We expect an increase in the maximum of the disk temperature as the R$_{\mathrm{FeII}}$ increases. According to Figure 1 from \cite{sh14},  increase in R$_{\mathrm{FeII}}$ implies increase in the Eddington ratio or decrease in the mass of the black hole. We expect that this maximum temperature depends not only on the Eddington ratio \citep{c06} but on the ratio of the Eddington ratio to the black hole mass (or, equivalently, on the ratio of the accretion rate to square of the black hole mass).

\section{Theory}


Most of the quasar radiation comes from the accretion disk and forms the Big Blue Bump (BBB) in the optical-UV (\citealt{c87, rich06}), and this thermal emission is accompanied by an X-ray emission coming from a hot optically thin mostly compact plasma, frequently refered to as a corona \cite{c87,haa91,Fabian2015}. The ionizing continuum emission thus consists of two physically different spectral components. We parameterize this emission in the following way. For convenience, the BBB component is parameterized by the maximum temperature of an accretion disk. In the standard accretion disk model this temperature is related to the black hole mass and accretion rate 

\begin{equation}
\text{T}_{\text{BBB}} = \left[\frac{3\text{GM}\dot{\text{M}}}{8\pi \sigma \text{r}^3}\left(1 - \sqrt{\frac{\text{R}_{\text{in}}}{\text{r}}}\right)\right]^{0.25} = 1.732\times 10^{19} \left(\frac{\dot{\text{M}}}{\text{M}^{2}}\right)^{0.25}\label{eq:01}
\end{equation}
where $\text{T}_{\text{BBB}}$ - maximum temperature corresponding to the Big Blue Bump; G - gravitational constant; M - black hole mass; $\dot{\text{M}}$ - black hole accretion rate; r - radial distance from the centre; $\mathrm{R_{in}}$ - radius corresponding to the innermost stable circular orbit. $\text{M}$ and $\dot{\text{M}}$ are in cgs units. Similar formalism has been used by \cite{b07} although the coefficient differs by a factor of 2.6 from Equation 1. This maximum is achieved not at the innermost stable orbit around a non-rotating black hole (3R$_{\mathrm{Schw}}$) but at  4.08$\bar{3}$ R$_{\mathrm{Schw}}$.  The SED component peaks at the frequency
\begin{equation}
\nu_{\text{max}} \sim \left[\frac{\frac{\text{L}}{{\text{L}_{\text{Edd}}}}}{\text{M}}\right]^{\text{0.25}}\label{eq:02}.
\end{equation}
where $\nu_{\text{max}}$ - frequency corresponding to $\text{T}_{\text{BBB}}$; L - accretion luminosity $\left (=\eta \dot{\text{M}} \text{c}^2 \right )$; $L_{\text{Edd}}$ - Eddington limit $\left (= \mathrm{\frac{4\pi GMm_{p}c}{\sigma_{T}}}, \text{where\;} \mathrm{m_{p}} - \text{mass\; of\; a\; proton}, \sigma_{\text{T}} - \text{Thompson \;cross\; section}\right )$. The exact value of the proportionality coefficient has to be calculated numerically, and for a standard Shakura-Sunyaev disk $\mathrm{h \nu_{max}/k T_{BBB}} = 2.092$. We expect that the thin-disk formalism applies to all the Type 1 AGN radiating above 0.01$L_{\text{Edd}}$ and below 0.3$L_{\text{Edd}}$. Instead of a full numerical model of an accretion disk spectrum, we simply use a power law with the fixed slope, $\alpha_{uv}$, and the value of $\text{T}_{\text{BBB}}$ to determine an exponential cut-off.  The X-ray coronal component shape is defined by the slope ($\mathrm{\alpha_{x}}$) and has an X-ray cut-off. The relative contribution is determined by fixing the broad band spectral index $\mathrm{\alpha_{ox}}$, and finally the absolute normalization of the incident spectrum is set assuming the source bolometric luminosity. We fix most of the parameters, and $\text{T}_{\text{BBB}}$ is the the basic parameter of our model. 

Some of this radiation is reprocessed in the BLR which produces the emission lines. In order to calculate the emissivity, we need to assume the mean hydrogen density ($\mathrm{n_H}$) of the cloud, and a limiting column density (N$_\mathrm{H}$) to define the outer edge of the cloud. Ionization state of the clouds depends also on the distance of the BLR from the nucleus.  We fix it using the observational relation by \cite{b13}

\begin{equation}
\left(\frac{\text{R}_{\text{BLR}}}{1\;\text{lt-day}}\right) = 10^{\left[1.555 + 0.542\; \text{log}\left(\frac{\lambda \text{L}_{\lambda}}{10^{44} \;\text{erg s}^{-1}}\right)\right]}\label{eq:03}
\end{equation}

The values for the constants considered in Equation 3 are taken from the Clean $\mathrm{H \beta\; R_{BLR} - L}$ model from \citet{b13} where $\lambda$ = 5100 \AA.

\section{Results and Discussions}



\begin{figure}[h!]
\begin{center}
\includegraphics[height=17.5cm, width=10cm, angle=270]{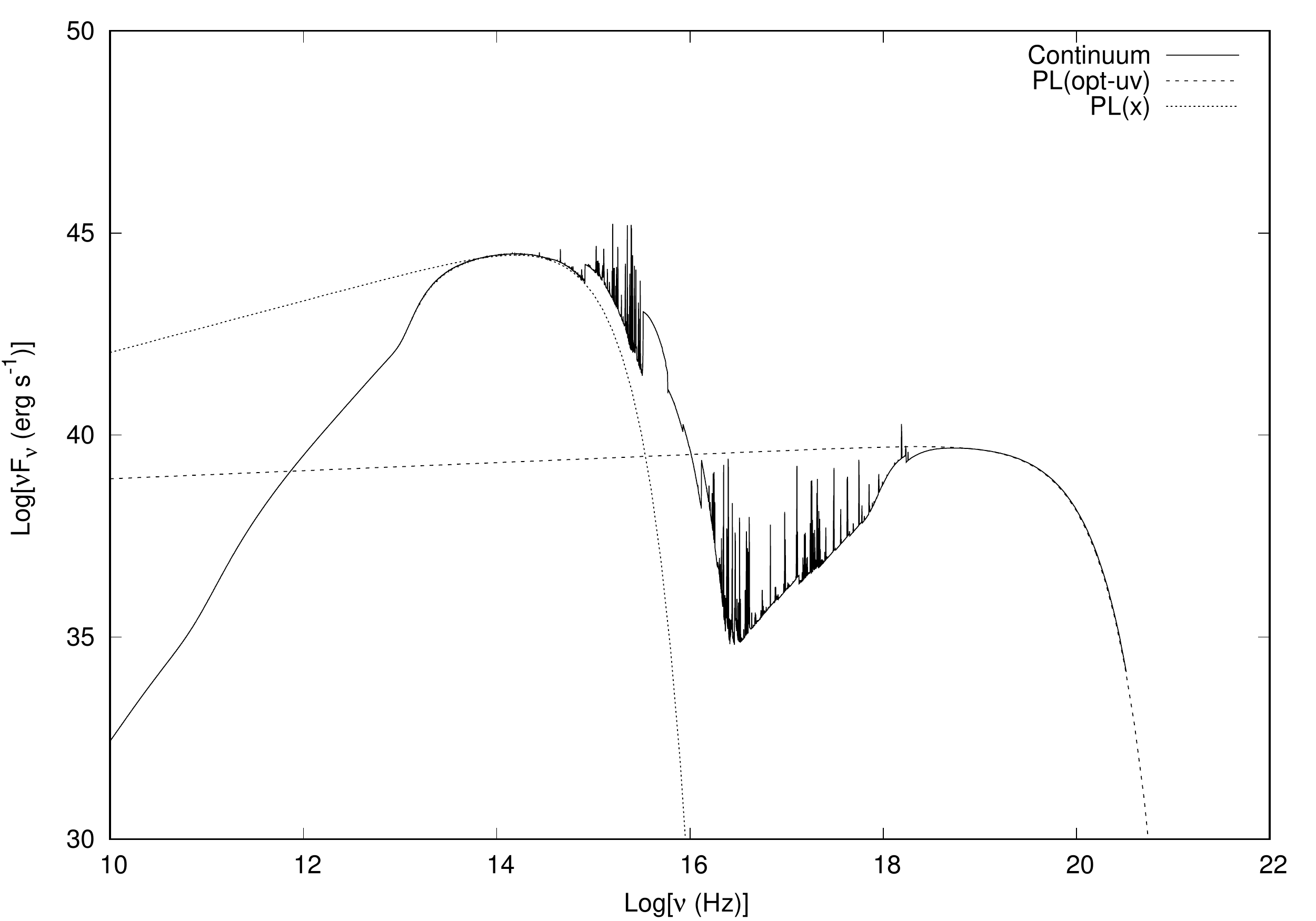}
\hfill
\hspace{0.25cm}
\includegraphics[height=8.5cm, width=17.5cm]{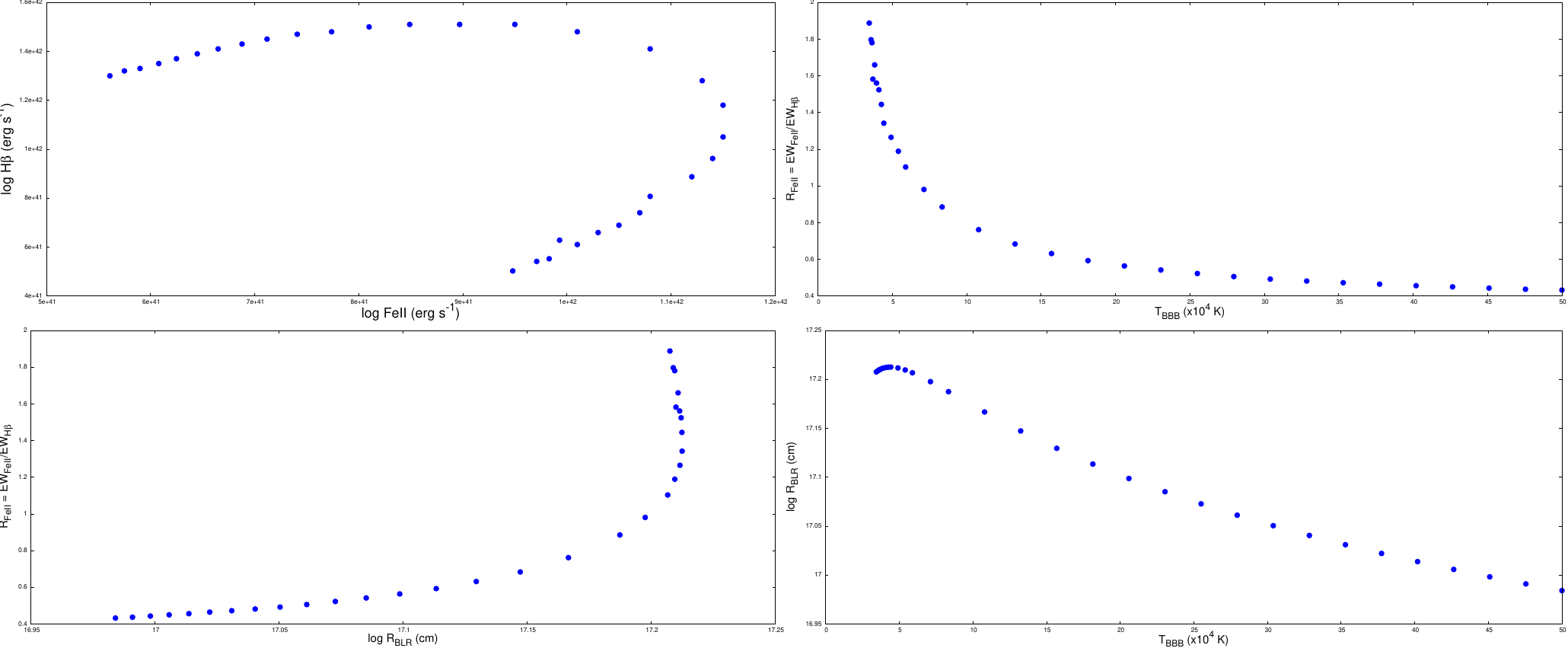}
\end{center}
\caption{\textbf{(A)} An example of the spectral energy distribution of the AGN transmitted continuum and line emission produced by CLOUDY: $\mathrm{T_{\mathrm{BBB}}} = 3.45\times 10^4 \;\mathrm{K}$, $\mathrm{\alpha_{\mathrm{ox}}} = -1.6$, $\mathrm{\alpha_{\mathrm{uv}}} = -0.36$, $\mathrm{\alpha_{\mathrm{x}}} = -0.9$,  $\mathrm{n_{\mathrm{H}}} = 10^{11} \;\mathrm{cm^{-3}}$, $\mathrm{N_{H}} = 10^{24} \;\mathrm{cm^{-2}}$ and $\mathrm{log\left(R_{BLR}\right)} = 17.208$. The two-component power law serves as the incident radiation; \textbf{(B)} The family of solutions is parameterized by maximum accretion disk temperature: (i) composite Fe$_{\mathrm{{II}}}$ line luminosity [$\mathrm{erg\;s^{-1}}$] - ${\mathrm{H\beta}}$ line luminosity [$\mathrm{erg\;
s^{-1}}$]; (ii) T$_{\mathrm{BBB}}$ - $\text{R}_{\mathrm{Fe{II}}}$; (iii) log R$_{\mathrm{BLR}}$ - $\text{R}_{\mathrm{Fe{II}}}$; (iv) T$_{\mathrm{BBB}}$ - log R$_{\mathrm{BLR}}$ . Trends plotted for the following values of the parameters: $\mathrm{L_{bol}} = 10^{45}\; \mathrm{erg\;s^{-1}}$, $\mathrm{n_{\mathrm{H}}}$ = $10^{11} \;\mathrm{cm^{-3}}$ and $\mathrm{N_{H}}$ = $10^{24} \;\mathrm{cm^{-2}}$.}\label{fig:SED}
\end{figure}

As a first test we check the dependence of the change in the R$_{\mathrm{FeII}}$ as a function of the accretion disk maximum temperature, T$_\mathrm{{BBB}}$  at constant values of L$_\mathrm{bol}$, $\mathrm{\alpha_{uv}}$ , $\mathrm{\alpha_{ox}}$ , $\mathrm{n_H}$ and $\mathrm{N_H}$. We fix the bolometric luminosity at the AGN, $\mathrm{L_{bol}}$ = $\mathrm{10^{45} \;erg\; s^{-1}}$ with accretion efficiency $\epsilon$ = 1/12, since we consider a non-rotating black hole in Newtonian approximation (see Eq.~1). This determines the accretion rate, $\dot M$. The BBB's exponential cutoff value is determined by the maximum temperature of the disk. Our branch of solutions covers the disk temperature range between $1.06\times 10^4$\;K and $1.53\times 10^5$\;K. The corresponding range of the black hole mass range obtained from Equation (1) is [$2.35\times 10^7\; \mathrm{M_{\odot}}$, $4.90\times 10^9\; \mathrm{M_{\odot}}$], and it implies the range of Eddington ratio ($\mathrm{L/L_{Edd}}$) [0.002, 0.33] calculated from the mentioned range of maximum disk temperatures. Large disk temperature corresponds to low black hole mass, since we fix the bolometric luminosity. Finally, we use a two-power law SED with optical-UV slope , $\mathrm{\alpha_{uv}}$ = -0.36, and X-ray slope, $\mathrm{\alpha_{x}}$ = -0.91 \citep{roz14}. The exponential cutoff for the X-ray component is fixed at 100 keV (\citealt{fra02} and references therein). By setting a value for the spectral index, $\mathrm{\alpha_{ox}}$ = -1.6, we specify the optical-UV and X-ray luminosities.  An example of SED is shown in upper panel of Figure~\ref{fig:SED}. 
\par

We now use this one-dimensional family of SED to calculate the line emission. We have dropped the X-ray power-law component in the subsequent analyses which we plan to re-introduce once we start to see the expected trend in the R$_{\mathrm{FeII}}$ - $\mathrm{T_{BBB}}$. As a start, we use the values of parameters from \cite{bv08} i.e. $\mathrm{log[n_H/(cm^{-3})] = 11}$, $\mathrm{log[N_H/(cm^{-2})] = 24}$, without including microturbulence (the motion that occurs within a cloud's line-forming region to whose variation the line formation and the emission spectrum is sensitive). The distance of the cloud from the source depends on adopted disk temperature. From the incident continuum, we estimate the $\mathrm{L_{5100 \AA}}$ that in turn is used to calculate the inner radius of the BLR cloud using Equation 3. 
\par

Knowing the irradiation,  we produce the intensities of the broad Fe${\mathrm{II}}$ emission lines from the corresponding levels of transitions present in CLOUDY 13.04 \citep{f13}. We calculate the Fe${\mathrm{II}}$ strength (R$_{\mathrm{FeII}}$ = EW$_{\mathrm{Fe{II}}}$ / EW$_{\mathrm{H\beta}}$), which is the ratio of Fe${\mathrm{II}}$ EW within 4434-4684 \AA \;to broad H$\beta$ EW. This prescription is taken from \cite{sh14}.
\par

The results are shown in lower panel of Figure~\ref{fig:SED}. The rise of the disk temperature initially leads to weak change of the cloud distance from the source, since the SED maximum is close to 5100 \AA~ for such massive black holes, and later with increasing distance from the source it decreases. The Fe${\mathrm{II}}$ intensity changes monotonically with $\mathrm{T_{BBB}}$ but it is a decreasing, not an increasing trend. This is not what we have expected --- high temperatures should correspond to low mass high accretion rate sources (\citealt{ss73}), Narrow Line Seyfert 1 galaxies, which show strong Fe${\mathrm{II}}$ component. This monotonic trend appears despite non-monotonic change with the disk temperature both in H$\beta$ and Fe${\mathrm{II}}$ itself.
\par

\begin{figure}[h!]
\begin{center}
\includegraphics[height=17.5cm, width=10cm, angle=270]{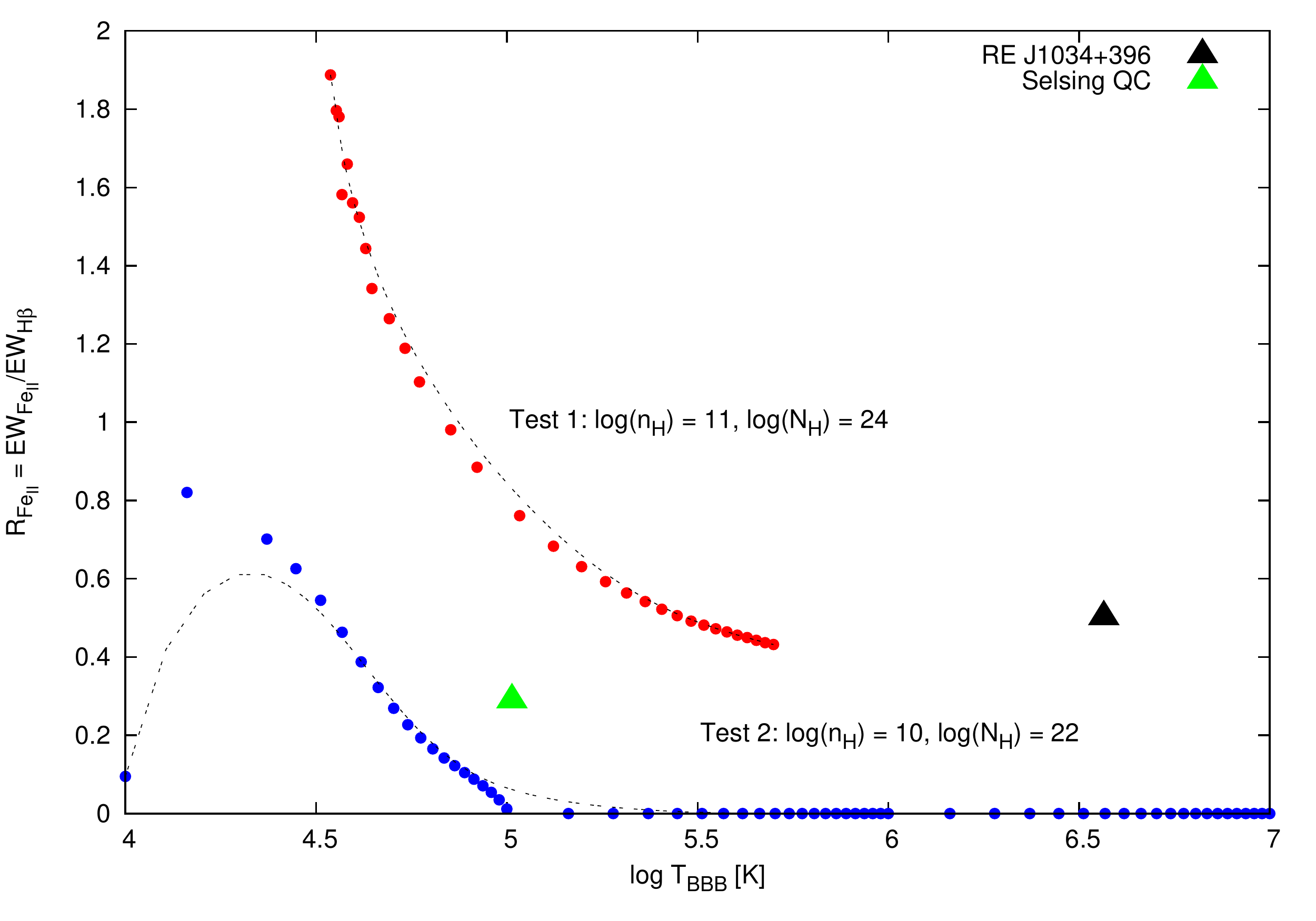}
\end{center}\caption{Comparison between  $\text{R}_{\mathrm{Fe{II}}}$ - T$_{\mathrm{BBB}}$ for two different constant-density single cloud models. Observational data RE J1034+396 (\citealt{c16}) \& Selsing quasar composite (\citealt{sel16}) plotted on the simulated trends. }\label{fig:R_Tbb_2}
\end{figure}

We thus extend our study for a broader parameter range, allowing for $\mathrm{log(n_H)}$ in the range 10 - 12, and $\mathrm{log(N_H)}$ from the range 22.0 - 24.0.  The range of values obtained for R$_{\mathrm{FeII}}$ went up from [0.005, 0.4] to [0.4, 1.95] with increasing $\mathrm{N_H}$. The change in the local density is also important. For a constant $\mathrm{log(N_H)} = 24$, changing $\mathrm{log(n_H)}$ = 10 - 12 shifts the maximum of R$_{\mathrm{FeII}}$ from 1.93 (for $\mathrm{log(n_H)}$ = 10) down to 0.095 (for $\mathrm{log(n_H)}$ = 12), thus, there is a declining trend in the maximum of R$_{\mathrm{FeII}}$ with an increase in $\mathrm{n_H}$ at constant $\mathrm{N_H}$. We see a definite change in the trend going from lower mean density to higher in the character of $\mathrm{T_{BBB} - {R_{Fe{II}}}}$ dependence. In the case of the lower $\mathrm{n_H}$ case, we see the turnover peak close to $\mathrm{log[T_{BBB}(K)] = 4.2}$ which couldn't be reproduced by the models generated using higher values of $\mathrm{n_H}$ and $\mathrm{N_H}$ owing to non-convergence of the CLOUDY code at lower values of T$_{\mathrm{BBB}}$. But on the higher end of T$_{\mathrm{BBB}}$ we still get the same declining behaviour of R$_{\mathrm{FeII}}$. The two extreme cases of changing both parameters are in Figure~\ref{fig:R_Tbb_2}. We thus find that the obtained values of R$_{\mathrm{FeII}}$ are heavily affected by the change in the maximum temperature of the BBB-component. The range of the R$_{\mathrm{FeII}}$ is well covered, in comparison with the plots of \cite{sh14}: higher density solutions reproduce large values and lower values are obtained by lowering the local density and column density. But, in general, there is a decay in the $\mathrm{Fe{II}}$ strength with the rise of the disk temperature while it was expected to follow a rising curve. 
\par

\begin{figure}[h!]
\begin{center}
\includegraphics[height=15cm,width=22.5cm, angle = 90]{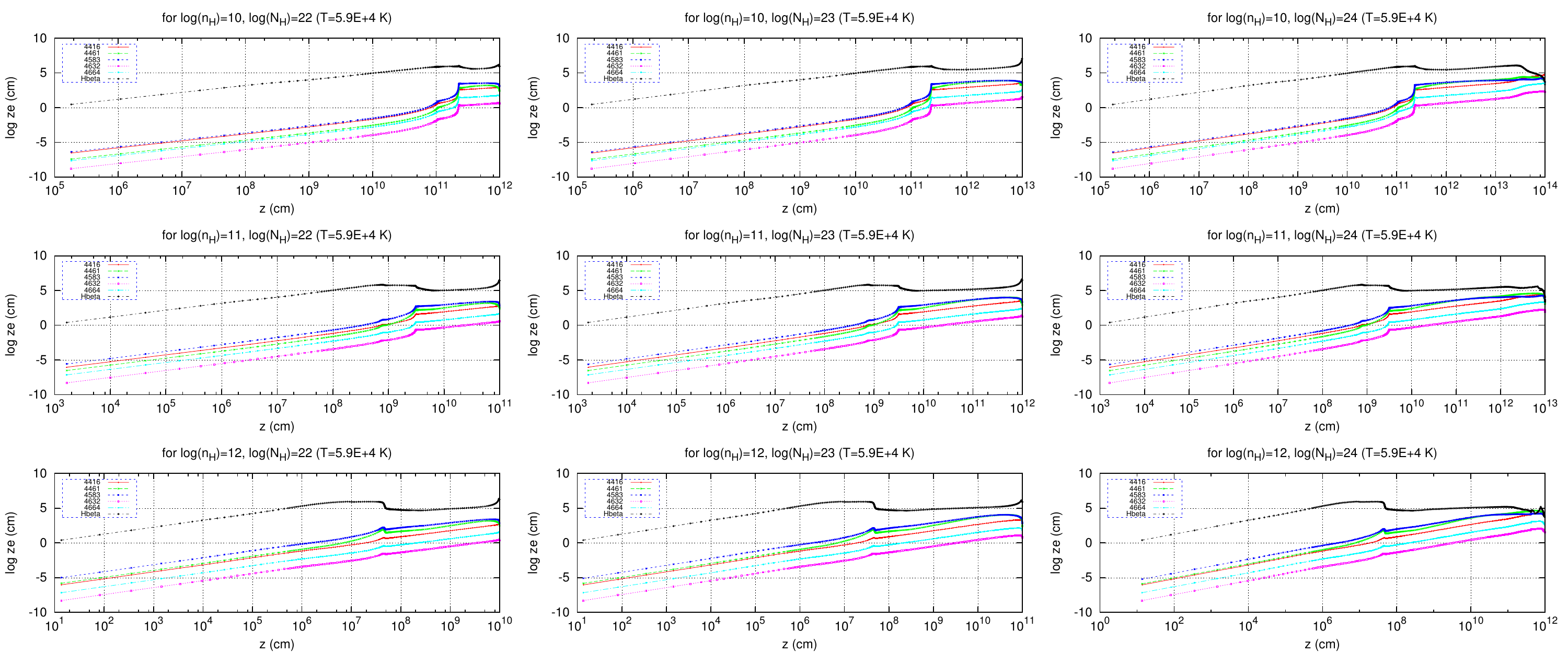}
\end{center}
\caption{Emissivity profiles of H$\mathrm{\beta}$ and Fe${\mathrm{II}}$ for different local densities and column densities: T$_{\mathrm{BBB}}$ = $5.9\times 10^{4}$ K. X-axis label: geometrical depth (in cm); Y-axis label: geometrical depth (z) $\times$ emissivity (e). }\label{fig:prof1}
\end{figure}

\begin{figure}[h!]
\begin{center}
\includegraphics[height=15cm,width=22.5cm, angle = 90]{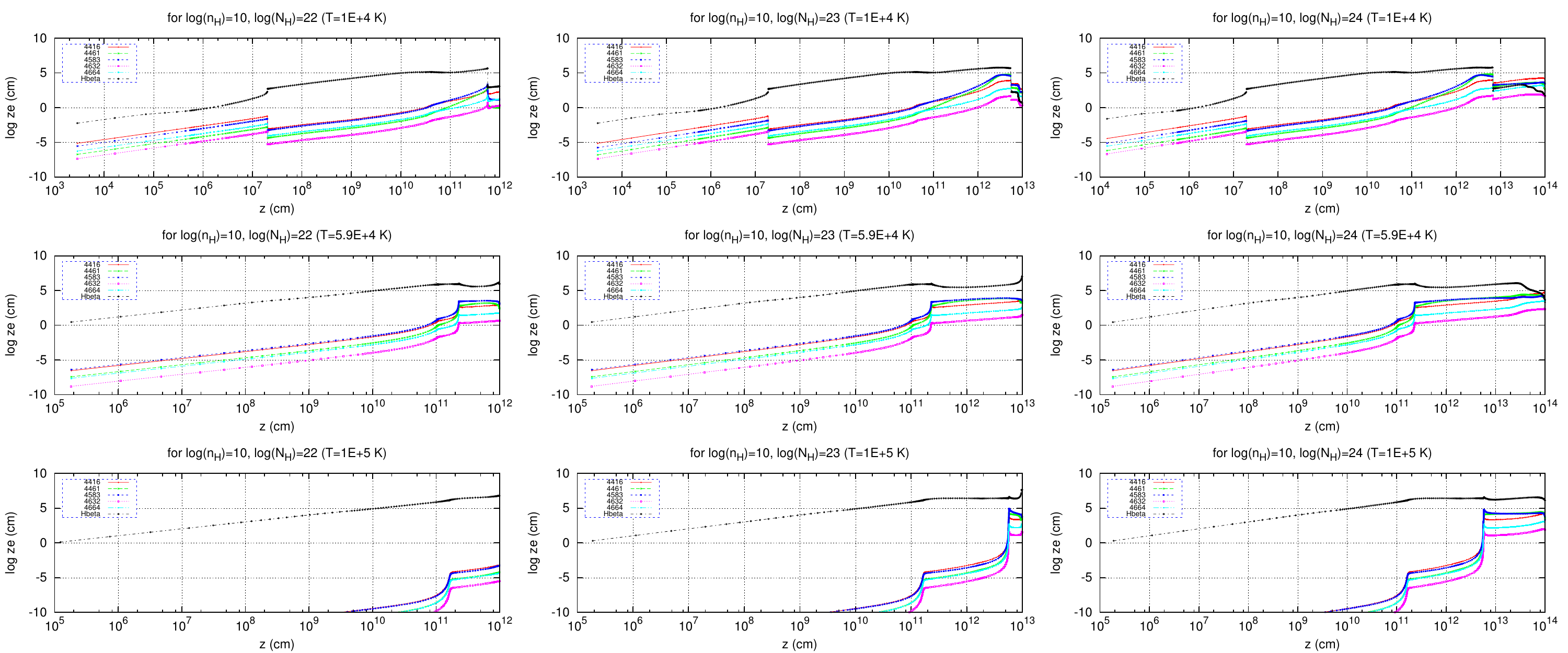}
\end{center}
\caption{Emissivity profiles of H$\mathrm{\beta}$ and Fe${\mathrm{II}}$: comparison at three different T$_{\mathrm{BBB}}$ values and column densities. X-axis label: geometrical depth (in cm); Y-axis label: geometrical depth (z) $\times$ emissivity (e).}\label{fig:prof2}
\end{figure}

To understand the nature of this trend in our CLOUDY computations we plot H$\beta$ and Fe${\mathrm{II}}$  emissivity profiles (Figure~\ref{fig:prof1} and ~\ref{fig:prof2}) where we consider only the first five Fe${\mathrm{II}}$ transitions in the 4434-4684 \AA\;range. We compute these profiles by varying $\mathrm{n_H}$ ($\mathrm{log(n_H)}$ = [10,12]), $\mathrm{N_H}$ ($\mathrm{log(N_H)}$ = [22,24]) and testing the dependence of $\mathrm{T_{BBB}}$ for three different temperature cases. The H$\beta$ nearly always dominates over the selected Fe${\mathrm{II}}$ emissions. But close to the outer surface of the cloud i.e., as $\mathrm{log(N_H) \rightarrow 24}$, the H$\beta$ emission starts to drop while the Fe${\mathrm{II}}$ increases with increasing $\mathrm{N_H}$, and there is some overlap region (see Figure~\ref{fig:prof1} and ~\ref{fig:prof2}). In Fig. ~\ref{fig:prof1}, we find that with increasing $\mathrm{n_H}$, the peak of the H$\beta$ formation shifts closer to the inner surface of the cloud, so the relative contribution of Fe${\mathrm{II}}$ rises. However, with increasing $\mathrm{T_{BBB}}$ the emissivity zones move deeper, and the relative role of H$\beta$ (see the extreme right panel of Fig. ~\ref{fig:prof2}) goes down.
\par

In general, the emissivity profile is much more shallow for H$\beta$ while Fe${\mathrm{II}}$ emission is more concentrated towards the back of the cloud. Thus, an increase in $\mathrm{N_H}$ brings the R$_{\mathrm{FeII}}$ ratio up, but increasing irradiation pushes the H$\beta$ and Fe$\text{II}$ emitting regions deeper into the cloud and  R$_{\mathrm{FeII}}$ drops (see Figure~\ref{fig:prof1} and ~\ref{fig:prof2}). Our sequence of solutions for fixed bolometric luminosity and rising accretion disk temperature creates an increasing irradiation, and apparently the change of the SED shape cannot reverse the trend. 
\par

Therefore, the question is whether our hypothesis of the dominant role is incorrect or the set of computations is not satisfactory. To answer it we used two objects with well measured SED as well as R$_{\mathrm{FeII}}$: RE J1034+396 \citealt{c16} and an X-Shooter quasar composite from \citealt{sel16}. In order to determine the parameter $\text{T}_{\text{BBB}}$ for those sources we created a set of full-GR disk models following the Novikov-Thorne prescription, we simulate an array of SED curves with $\mathrm{L_{Edd}}$ parametrization where we consider simultaneous dependence on spin ($0 \le \mathrm{a} \le 0.998$) and accretion rate ($0.01 \le \mathrm{\dot{m}} \le 10$; where $\mathrm{\dot{m}} = \mathrm{\dot{M}/{\dot{M}_{Edd}}}$, $\mathrm{\dot{M}} = 1.678\times 10^{18}\mathrm{\frac{M}{M_{\odot}}}$). The value of the black hole mass has been taken from \cite{cap16}. The two values represent the extreme tails of the possible trend, with X-Shooter composite having the SED peak in UV and RE J1034+396 peaikng in soft X-rays. The corresponding points are shown in  Figure~\ref{fig:R_Tbb_2}. Observations show a rise in the value of R$_{\mathrm{FeII}}$ with increase in $\mathrm{T_{BBB}}$ which the simulations have been unable to reproduce so far. However, the rise in  R$_{\mathrm{FeII}}$ is not very large, from 0.3 to 0.5, despite huge change in the disk temperature difference implied by the observed SED in the two objects.

\section{Future}
The reason for starting the project from purely theoretical modeling of the line ratios is the fact that determinations of the black hole mass, accretion rate and the observational parameter R$_{\mathrm{FeII}}$ available in the literature are not accurate enough to be used to test our hypothesis about the nature of the EV1 \citep{s17}.  The subsquent tasks will be to check the R$_{\mathrm{FeII}}$ dependence on other parameters (see \citealt{sul00, mao09, marz15}) which are used (L$_\mathrm{bol}$ , $\mathrm{\alpha_{uv}}$ , $\mathrm{\alpha_{ox}}$ , $\mathrm{n_H}$, $\mathrm{N_H}$, $\mathrm{cos}(i)$, $\mathrm{a}$ and others). We intend to incorporate the microturbulence as suggested in \cite{bv08}. Better, more physical description of the SED may be needed, i.e  model of a disk + corona with full GR and more complex geometry of the BLR using \cite{c11}, \cite{ch11}, and \cite{c15}. We intend to implement the constant density LOC (Locally Optimized Cloud) model and subsequently the constant pressure model to repeat the tests and check for discrepancies with respect to the current model. Next stage to consider is the possibility of shielding of some BLR regions by the puffed inner disk \cite[e.g.][]{Wang2014}, or to consider independent production regions of  H$\beta$ and Fe${\mathrm{II}}$.  Different studies have proposed that Fe${\mathrm{II}}$ is mainly produced in BLR (\citealt{bv08,shi10}) while many others have suggested that these emissions are mostly produced in the accretion disk (\citealt{mar15} and references therein). Finally, we have to test our theory observationally for more sources with known SED peak position. To have an overview of the EV1, it is necessary to study it in other ranges of frequencies, including X-ray, radio, UV and IR spectral ranges. Considerable progress along these line have been made by  \cite{Sul2007}, \cite{Sul17} (UV range), \cite{dul99} (Figure 4), \cite{mar15} (IR range).

\section*{Conflict of Interest Statement}

The authors declare that the research was conducted in the absence of any commercial or financial relationships that could be construed as a potential conflict of interest.

\section*{Author Contributions}
SP has tested the basic model and carried out the photoionisation simulations based on the idea and formalism proposed by BC. CW has provided computational assistance and helped solve the $\mathrm{T_{BBB}}$ issue.

\section*{Funding}
Part of this work was supported by Polish grant Nr. 2015/17/B/ST9/03436/.

\section*{Acknowledgments}
The authors would like to acknoweledge the anonymous referees for their comments and suggestions to bring the paper to its current state. SP would like to acknowledge the organising committee and the participants of the “Quasars at all Cosmic Epochs” conference held during 02nd Apr - 07th Apr 2017 in Padova, Italy and, subsequently for providing the opportunity to present a talk on his research after being adjudged with the Best Poster award. SP would also like to extend his gratitude to the Center for Theoretical Physics and Nicolaus Copernicus Astronomical Center, Warsaw, the National Science Center (NSC) OPUS 9 grant for financing the project and Dr. Gary Ferland and Co. for the photoionisation code CLOUDY. SP would like to acknowledge the unending academic and personal support from Mr. Tek Prasad Adhikari. SP is also obliged to the Strong Gravity group at CAMK, Warsaw for engaging discussions resulting in this work.


\bibliographystyle{frontiersinSCNS_ENG_HUMS} 
\bibliography{frontiers}

\newpage
\section*{Figure captions}


\vspace{-10cm}




\end{document}